\documentclass[12pt,a4paper]{article}
\pdfoutput=1
\usepackage[a4paper]{geometry}

\usepackage{amsmath}
\usepackage{amsfonts}
\usepackage{amssymb}
\usepackage{dsfont}
\usepackage[nosort]{cite}
\usepackage{hyperref}
\usepackage{multirow}

\newcommand{\eq}[1]{\begin{equation}
                     \begin{split} #1 \end{split}
                     \end{equation}}
\newcommand{\ov}{\overline}
\newcommand{\op}{\hspace{1pt}}

\allowdisplaybreaks[2]
\numberwithin{equation}{section}

\begin{document}

\vspace*{-1.5cm}
\begin{flushright}
  {\small
  LMU-ASC 04/20
  }
\end{flushright}

\vspace{1.75cm}

\begin{center}
{\LARGE
Swampland conjectures for type IIB 
orientifolds \\ 
with closed-string $U(1)$s \\
}
\end{center}

\vspace{0.4cm}

\begin{center}
  Mart\'in Enr\'iquez Rojo, Erik Plauschinn
\end{center}

\vspace{0.3cm}

\begin{center} 
\textit{Arnold Sommerfeld Center for Theoretical Physics\\[1pt]
Ludwig-Maximilians-Universit\"at \\[1pt]
Theresienstra\ss e 37 \\[1pt]
80333 M\"unchen, Germany}
\end{center} 

\vspace{1.8cm}


\begin{abstract}
\noindent
We study the weak gravity conjecture, 
the swampland distance conjecture and the emergence 
proposal for $\mathcal{N}=1$ orientifold 
compactifications of type IIB string theory with 
O3-/O7-planes. We allow for orientifold projections 
with $h^{2,1}_+\neq 0$ which gives rise to
closed-string $U(1)$ gauge fields, and our
findings show that certain structures present for
$\mathcal N=2$ compactifications are not present 
for $\mathcal N=1$. In particular, 
assumptions about stability have to be relaxed
and we encounter an ambiguity for the emergence
of gauge symmetries associated with
the $h^{2,1}_+$ sector.
\end{abstract}


\clearpage

\tableofcontents


\section{Introduction}

String theory is argued to be 
a consistent theory of quantum gravity 
including gauge interactions. 
As such, it provides an ideal testing 
ground for expectations 
a general quantum-gravity theory should satisfy.
However, conversely one can try to identify 
quantum-gravity features in string theory
and conjecture them to hold in general
--- this is the approach known as the swampland program.

The swampland is defined as the set
of consistent 
effective quantum field theories which cannot be completed 
into consistent theories of quantum gravity
\cite{Vafa:2005ui}. During the last years there
has been an immense effort in pursuing 
the swampland program and in proposing 
and testing swampland conjectures.
It is not possible to give an overview of 
the corresponding literature here, and
we therefore want to 
refer to the review article
\cite{Palti:2019pca}. 
There are however three conjectures
which play a role for this paper:
these are 
 the weak gravity conjecture in its original 
form \cite{ArkaniHamed:2006dz} 
and the weak gravity conjecture with scalar fields
\cite{Palti:2017elp};
 the swampland distance conjecture
\cite{Ooguri:2006in} in its refined version
\cite{Klaewer:2016kiy};
and the emergence proposal 
\cite{Grimm:2018ohb} which has been 
further developed in \cite{Palti:2019pca}.
These conjectures will be stated in 
section~\ref{sect_swampconj}.

The swampland program typically does not 
rely on supersymmetry, but many 
 computations in string theory 
are simplified considerably when 
some amount of supersymmetry is preserved. 
A well-established setting is that of 
type IIB string theory compactified on Calabi-Yau 
three-folds $\mathcal X$, for which
the resulting 
theory in four dimensions is $\mathcal N=2$ 
supersymmetric. The massless spectrum 
is determined by the topology of the 
compact space, in particular, the 
four-dimensional theory contains
one gravity multiplet, $h^{2,1}$ vector multiplets
and $h^{1,1}+1$ hypermultiplets where 
$h^{2,1}$ and $h^{1,1}$ denote the non-trivial 
Hodge numbers of $\mathcal X$.
In such compactifications one
obtains a variety of BPS objects which have 
special properties and features, 
given for instance by D-branes wrapping 
cycles of the compact space.
The authors of \cite{Grimm:2018ohb}
consider such type IIB compactifications
and focus on the vector-multiplet sector, which
contains closed-string $U(1)$ gauge fields and the
complex-structure moduli of $\mathcal X$. 
The relevant BPS objects are D3-branes
wrapping three-cycles of the Calabi-Yau manifold, 
and they are charged under the 
$U(1)$s and their mass depends
on the complex-structure moduli. 
For this setting many explicit checks 
of various swampland conjectures have been 
performed and the inter-dependence of 
some of these conjectures has been illustrated.

The main question we want to ask in 
this paper is 
\textit{What happens to the $\mathcal N=2$ 
analysis when we perform 
an orientifold projection?}
It is well-known that under an 
orientifold projection (giving rise to O3- and O7-planes)
the $h^{2,1}$ vector multiplets of 
$\mathcal N=2$ are projected to 
$h^{2,1}_-$ chiral multiplets
and $h^{2,1}_+$ vector multiplets of 
four-dimensional $\mathcal N=1$ supergravity
\cite{Grimm:2004uq}. In particular, we have
\eq{
  \label{multiplets_split}
  \renewcommand{\arraystretch}{1.6}
  \arraycolsep15pt
  \begin{array}{ccl}
  \mathcal N=2 && 
  \hfill\mathcal N=1\hfill
  \\
  \hline
  \multirow{2}{*}{\mbox{$h^{2,1}$ vector multiplets}}
  &
  \multirow{2}{*}{$\xrightarrow{\hspace{5pt}{\rm orientifold}\hspace{5pt}}$}
  & \mbox{$h^{2,1}_-$ chiral multiplets}
  \\
  && \mbox{$h^{2,1}_+$ vector multiplets}
  \end{array}
}
where $h^{2,1}=h^{2,1}_-+h^{2,1}_+$. Note that the $\mathcal N=1$ chiral multiplets 
contain the complex-structure moduli
and the $\mathcal N=1$ vector multiplets
contain the $U(1)$ gauge fields, and 
that both multiplets are independent of 
each other. 
As we will argue, this implies that 
D3-branes wrapping three-cycles are 
no longer BPS but split into massive-uncharged and massless-charged states in the four-dimensional 
theory.
Hence, a priori the  analysis of \cite{Grimm:2018ohb} no longer applies.

We now want to briefly mention some works in the 
literature which are  related to 
our discussion:
\begin{itemize}

\item As discussed above, the work 
of \cite{Grimm:2018ohb} studies swampland
conjectures for $\mathcal N=2$ compactifications
of type IIB string theory. 
Here we perform an analysis for the 
theory obtained after an orientifold projection 
to $\mathcal N=1$.

\item In \cite{Font:2019cxq} the authors
investigate (among others)
domain walls in $\mathcal N=1$ 
orientifold compactifications of 
type II string theory. For the O3-/O7-situation, these domain walls are given by D5-branes 
wrapping three-cycles of the Calabi-Yau manifold.
The mass of these four-dimensional states
is determined by the complex-structure 
moduli and they are charged under 
closed-string three-form gauge fields, but not 
under closed-string one-form gauge fields. 

\end{itemize}

This paper is organized as follows: in section~\ref{sec_iiborientifolds} we review type IIB orientifold compactifications with 
orientifold three- and seven-planes, and 
in section~\ref{sec_d3} we analyze 
D3-branes wrapping three-cycles in the 
compact space. In section~\ref{sect_swampconj} we discuss the weak gravity conjecture, the 
swampland distance conjecture and the 
emergence proposal for this setting,
and section~\ref{sect_conc} contains 
our summary and conclusions.
We furthermore mention that this paper is based on the master thesis \cite{martin:2019}
of one of the authors, 
where further details and discussions can 
be found.


\section{Type IIB orientifolds}
\label{sec_iiborientifolds}

In this section we briefly review Calabi-Yau orientifold compactifications of type IIB string theory with O3- and O7-planes. We consider the general situation with $h^{2,1}_+\neq0$,
leading to closed-string $U(1)$ gauge fields
in the four-dimensional effective theory \cite{Grimm:2004uq}.


\subsection{Calabi-Yau orientifolds}
\label{sec_orientifolds}

We start by introducing our notation for the topology of Calabi-Yau orientifolds,
and we provide some details on the special geometry of the complex-structure moduli space. 


\subsubsection*{Topology}

We consider a Calabi-Yau three-fold $\mathcal X$,
which comes with a  holomorphic three-form $\Omega\in H^{3,0}(\mathcal X)$ and a 
real K\"ahler form $J\in H^{1,1}(\mathcal X)$.
To perform an orientifold projection we impose 
a holomorphic involution 
$\sigma$ on $\mathcal X$, chosen to act on $\Omega$ and $J$ as
\eq{
  \label{orient_choice}
  \sigma^*J=+J\,,\hspace{60pt}
  \sigma^*\op\Omega=-\Omega\,.
}
Since $\sigma$ is an involution,
the cohomology groups of the Calabi-Yau three-fold $\mathcal X$ split into even and odd 
eigenspaces as $H^{p,q}(\mathcal X) = H^{p,q}_+(\mathcal X) \oplus 
H^{p,q}_-(\mathcal X)$ with dimensions $h^{p,q}_+$ and $h^{p,q}_-$.
Of interest to us will be the third de-Rham cohomology group 
$H^3(\mathcal X)= H^{3}_+(\mathcal X) \oplus 
H^{3}_-(\mathcal X)$, for which we choose a symplectic basis in the following way
\begin{align}
  &\label{basis_01}
  \arraycolsep2pt
  \begin{array}{l@{\hspace{40pt}}lcl@{\hspace{40pt}}lcl}
  \raisebox{-20pt}[0pt][-20pt]{$\displaystyle \{\alpha_I ,\alpha_a, \beta^I,\beta^a \}\,,$}
  &
  \displaystyle \int_{\mathcal X} \alpha_{I} \wedge \beta^J &=& \displaystyle \delta_I{}^J \,,
  &
  \displaystyle I,J &=& \displaystyle 0,\ldots, h^{2,1}_- \,,
  \\[16pt]
  &
  \displaystyle \int_{\mathcal X} \alpha_{a} \wedge \beta^b &=& \displaystyle \delta_a{}^b \,,
  &
 \displaystyle a,b &=& \displaystyle 1,\ldots, h^{2,1}_+ \,,
 \end{array}  
 \\
\intertext{
with all other pairings vanishing. 
A corresponding basis of the third homology group 
$H_3(\mathcal X)= H_{3+}(\mathcal X) \oplus 
H_{3-}(\mathcal X)$
together with the non-trivial pairings takes the form
}
  &\label{basis_02}
  \arraycolsep2pt
  \begin{array}{l@{\hspace{34pt}}lcl@{\hspace{62pt}}lcl}
  \raisebox{-20pt}[0pt][-20pt]{$\displaystyle \{A^I ,A^a, B_I,B_a \}\,,$}
  &
  \displaystyle \int_{A^I} \alpha_{J} &=& \displaystyle \delta^I{}_J \,,
  &
  \displaystyle \int_{B_I} \beta^{J} &=& \displaystyle \delta_I{}^J \,,
  \\[16pt]
  &
  \displaystyle \int_{A^a} \alpha_{b} &=& \displaystyle \delta^a{}_b \,,
  &
  \displaystyle \int_{B_a} \beta^{b} &=& \displaystyle \delta_a{}^b \,.
  \end{array}  
\end{align}


\subsubsection*{Special geometry}

Next, we consider the special geometry of the complex-structure moduli space. 
The holomorphic three-form $\Omega$ can be expanded in the symplectic basis
\eqref{basis_01}
as follows
\eq{
  \Omega = X^I \alpha_I - F_I\op \beta^I \,.
}
Since $\Omega$ is odd under the orientifold projection -- cf.~equation \eqref{orient_choice} -- the coefficients of $\Omega$
in the even basis $\{\alpha_a,\beta^a\}$ are vanishing, that 
is $X^a=0$ and $F_a\rvert_{X^a=0}=0$. We will refer to these conditions also as the orientifold locus
and discuss them in more detail below. The complex-structure moduli are coordinates on a projective space and are usually defined
as
\eq{
  z^i = \frac{X^i}{X^0} 
  \,, \hspace{40pt} z^i= u^i + i\op v^i
  \,, \hspace{40pt} i = 1, \ldots, h^{2,1}_- \,,
}
and we note that in our conventions the large-complex-structure limit corresponds to $v^i\to\infty$. 
For Calabi-Yau compactifications the periods $F_I$ can be defined in terms of a prepotential $F$ which is a
holomorphic function of degree two in the fields $X^{\mathsf I} = (X^I,X^a)$. 
More concretely, the $F_I$ are defined as the derivatives
\eq{
  F_I = \frac{\partial F}{\partial X^I} \biggr\rvert_{X^a=F_a=0} \,,
}
which in our present setting have to be evaluated at the orientifold locus. 
The prepotential $F$ of the $\mathcal N=2$ theory  consists of a tree-level part involving the triple intersection numbers
$d_{\mathsf{ijk}}$ of the mirror Calabi-Yau manifold and one-loop and non-perturbative 
corrections. More concretely, we have \cite{Hosono:1994av,Hosono:1994ax}
\eq{
\label{prepotential_full}
F=
\arraycolsep2pt
\begin{array}[t]{cl}
&\displaystyle \frac{d_{\mathsf{ijk}} X^{\mathsf i}X^{\mathsf j}X^{\mathsf k}}{X^{0}}
\\[8pt]
+&\displaystyle  \frac{1}{2}\op a_{\mathsf {ij}}X^{\mathsf i}X^{\mathsf j}+b_{\mathsf i}\op X^{\mathsf i}X^{0}
+\frac{1}{2}\op c\op (X^{0})^2
+ i\op (X^{0})^2\sum_{k}n_{k}\text{Li}_{3}\Bigl(e^{2\pi i\op k_{\mathsf i}X^{\mathsf i}/X^{0}}\Bigr) \,,
\end{array}
}
where the indices $\mathsf{i},\mathsf{j},\mathsf{k}=1,\ldots, h^{2,1}$ extend over the even and odd cohomology, 
$a_{\mathsf{ij}}$ and $b_{\mathsf i}$ are rational real numbers whereas $c$ is purely imaginary.\footnote{These numbers encode the one-loop corrections. 
We note that there is an ambiguity related to them:
for certain non-singular cases they have been 
computed in \cite{Hosono:1994ax} but
to our knowledge these are not known
for singular configurations.\label{foot_oneloop}}
The triple intersection numbers $d_{\mathsf{ijk}}$ and the $a_{\mathsf{ij}}$ are symmetric in their indices, 
$n_{k}$ are the genus zero Gopakumar-Vafa invariants, ${\rm Li}_3$ denotes the third poly-logarithm and
$k_{\mathsf i}$ are the wrapping numbers of 
world-sheet instantons along two-cycles of 
the mirror Calabi-Yau.


\subsubsection*{Orientifold locus}
\label{page_or_loc}

From the $\mathcal N=2$ perspective the orientifold locus 
is determined by $X^a=0$, $F_a=0$ for all $a=1,\ldots, h^{2,1}_+$.
The condition $X^a$ alone describes 
a conifold singularity in which an $A$-cycle
shrinks to zero size, and the physics 
of conifold singularities 
is well understood \cite{Strominger:1995cz}.
However, imposing in addition $F_a=0$ 
means that also the dual $B$-cycle vanishes.
We are not aware of work in the 
literature studying such situations and we 
continue with a heuristic discussion.

Let us then study in more 
detail the condition
$F_a=0$ evaluated at $X^a=0$. 
From the explicit form of
\eqref{prepotential_full} we determine 
\eq{
  \label{period_ll}
  F_a 
   = X^0 \left[ 3 \op d_{aij} z^i z^j
  + a_{ai} z^i + b_a
  - 2\pi 
  \sum_{k} k_a n_{k}\text{Li}_{2}\Bigl(e^{2\pi i\op k_{i}z^{i}}\Bigr)
  \right],
}
where we recall that $i,j=1,\ldots, h^{2,1}_-$ and 
$a=1,\ldots, h^{2,1}_+$.
Requiring $F_a=0$ for all $z^i$ 
can be solved in the following two ways:
\begin{enumerate}
\label{page_or_sol}

\item A sufficient set of conditions for 
the periods $F_a$  to vanish for all $z^i$
reads 
\eq{
  \label{or_sol_1}
  d_{aij} = 0 \,, \hspace{40pt}
  a_{ai} = 0 \,, \hspace{40pt}
  b_{a} = 0 \,, \hspace{40pt}
  k_{a} = 0 \,.
}
However, especially the condition $k_a=0$ is 
rather restrictive since it means that 
all world-sheet instanton corrections in 
the orientifold-even sector have to be absent.
This in turn implies a discrepancy 
in our discussion of emergence in section~\ref{emergencesect}.

\item A second possibility is to require a
splitting of the 
sum over $k$ in \eqref{period_ll} into 
purely odd and purely even vectors as
$
  \sum_{k} = \sum_{k_i} + \sum_{k_a}
$.
Using then the relation 
$\text{Li}_2(1) = \pi^2/6$ we find
the conditions
\eq{
  \label{or_sol_2}
  d_{aij} = 0 \,, \hspace{40pt}
  a_{ai} = 0 \,, \hspace{40pt}
  b_{a} = \frac{\pi^3}{3} \sum_{k_a} k_a n_k \,.
}
It is not clear to us whether the last of 
these conditions can be satisfied, 
but we mentioned in footnote~\ref{foot_oneloop} that there is an ambiguity for the 
one-loop coefficients. 
The solution \eqref{or_sol_2} however
realizes $F_a=0$ with  $k_a\neq0$
which is needed 
for our discussion in section~\ref{emergencesect}.

\end{enumerate}


\subsection{Effective action}

We now turn to the effective four-dimensional action obtained after compactifying 
type IIB string theory on the Calabi-Yau orientifolds described above. 
The details of this action can be found in the literature,
and here we focus on the sector corresponding to the third cohomology of $\mathcal X$.


\subsubsection*{Moduli}

We compactify type IIB string theory on a Calabi-Yau three-fold
as $\mathbb R^{3,1}\times \mathcal X$ and perform an orientifold projection.
The orientifold projection is of the form
$\Omega_{\rm P} (-1)^{F_{\rm L}} \sigma$, where 
$\Omega_{\rm P}$ denotes the world-sheet parity operator and $F_{\rm L}$ is 
the left-moving fermion number.
The holomorphic involution $\sigma$ acts on $\mathcal X$ as in 
\eqref{orient_choice} and 
leaves the non-compact four-di\-men\-sional part invariant. Hence,
the fixed loci of $\sigma$ correspond to orientifold three- and seven-planes.
The combination $\Omega_{\rm P} (-1)^{F_{\rm L}}$ acts on the massless bosonic fields 
in the following way
\eq{
  \label{orient_signs}
  \arraycolsep2pt
  \begin{array}{lclclcl}
  \arraycolsep1.5pt
  \displaystyle \Omega_{\rm P}\, (-1)^{F_{\rm L}}\: g &=& + \:g \,,  & \hspace{60pt} &
  \displaystyle \Omega_{\rm P}\, (-1)^{F_{\rm L}}\: B &=& - B \,, \\[6pt]
  \displaystyle \Omega_{\rm P}\, (-1)^{F_{\rm L}}\: \phi &=& + \:\phi \,,  & \qquad &
  \displaystyle \Omega_{\rm P}\, (-1)^{F_{\rm L}}\: C_{2p} &=& (-1)^{p} 
    \,C_{2p} \,,
  \end{array}
}
with $g$ the metric, $B$ the Kalb-Ramond field, $\phi$ the dilaton and 
$C_{2p}$ the Ramond-Ramond (R-R) potentials in type IIB. 
The degrees of freedom contained in these fields
become fields in the compactified theory, in particular, the deformations of the 
Calabi-Yau metric are contained in the K\"ahler form $J$ and the holomorphic three-form $\Omega$.
For the Ramond-Ramond four-form potential we note that $C_4$ can be expanded  as
\begin{align}
  \label{expansion_001}
  C_4 = C_4^{(0,4)} +{} &C_4^{(1,3)} + C_4^{(2,2)} + C_4^{(4,0)} \,,
  \\[2pt]
  \nonumber
  &\hspace{4pt}\downarrow
  \\[2pt]
  \nonumber
  &C_4^{(1,3)} = U^a\wedge  \alpha _a + V_a\wedge \beta^a \,,
\end{align}
where in $C_4^{(m,n)}$ the superscript $m$ denotes the degree in four dimensions and $n$ the 
degree on the compact space $\mathcal X$.
Note furthermore that $U^a$ and $V_a$ correspond to $U(1)$ vector fields in four dimensions. 
The four-form \raisebox{0pt}[0pt][0pt]{$C_4^{(4,0)}$} in $\mathbb R^{3,1}$ is not dynamical, and 
due to the self-duality conditions imposed on $C_4$ half of the degrees of freedom in the expansion 
\eqref{expansion_001}
are removed. Here we choose to remove \raisebox{0pt}[0pt][0pt]{$C_4^{(2,2)}$} and $V_a$ but keep 
\raisebox{0pt}[0pt][0pt]{$C_4^{(0,4)}$} and $U^a$, respectively.
The four-dimensional massless fields (in addition to the four-dimensional metric $g_{\mu\nu}$) then take the following schematic
form
\eq{
  \renewcommand{\arraystretch}{1.2}
  \arraycolsep2pt
  \begin{array}{l@{\hspace{40pt}}lcl@{\hspace{10pt}}c@{\hspace{10pt}}l}
  \mbox{complex-structure moduli} & \multicolumn{3}{@{}l}{\Omega}   & \longleftrightarrow & H^{3}_-(\mathcal X)\,,
  \\
  \mbox{$U(1)$ vector fields} & \multicolumn{3}{@{}l}{C_4^{(1,3)}}  & \longleftrightarrow & H^{3}_+(\mathcal X)\,,
  \\[6pt]
  \mbox{K\"ahler moduli} & J &+& C_4^{(0,4)} & \longleftrightarrow & H^{2,2}_+(\mathcal X)\,,  
  \\
  \mbox{odd moduli} & B &+& C_2 & \longleftrightarrow & H^{1,1}_-(\mathcal X)\,,  
  \\
  \mbox{axio-dilaton} & \phi &+& C_0 & \longleftrightarrow & H^{0,0}_+(\mathcal X)\,.
  \end{array}
}


\subsubsection*{Four-dimensional action}

Compactifications of type II string theory on Calabi-Yau manifolds lead to a $\mathcal N=2$ 
supergravity theory in four dimensions, which can be
reduced to $\mathcal N=1$ by an orientifold projection. 
Here we are interested in the action corresponding to the complex-structure moduli $z^i$ and the vector fields 
$U^a$, which can be brought into the following  general form
\eq{
  \label{action_001}
  \mathcal S \supset
   \frac{1}{2}\int_{\mathbb R^{3,1}} \,\Bigl[ \;
  -\mbox{Re}\op (f_{ab}) \op F^a\wedge \star F^b
  +\mbox{Im}\op (f_{ab}) \op F^a\wedge  F^b    
  - 2\op G_{i\ov j} \, dz^i \wedge\star d\ov z{}^{\ov j} \;\Bigr]\,.
}
The K\"ahler metric $G_{i\ov j}$ for the complex-structure moduli $z^i$ appearing in the 
action \eqref{action_001} is computed from a K\"ahler potential $K_{\rm cs}$ as
\eq{
  \label{metric_k}
  G_{i\ov j} = \frac{\partial^2 K_{\rm cs}}{\partial z^i \partial \ov z{}^{\ov j}}\biggr\rvert_{X^c=F_c=0} \,,
  \hspace{50pt}
  K_{\rm cs} &= - \log \Bigl[\, - i\int_{\mathcal X} \Omega\wedge \ov\Omega\:\Bigr] \,,
}
and the $U(1)$ field strength corresponding to $U^a$ is denoted by $F^a = dU^a$.
The gauge kinetic function $f_{ab}$ can be expressed in terms of the period matrix
$\mathcal N$ as 
\eq{
  \label{period_m_001}
  &f_{ab} = -i \op\ov{\mathcal  N}_{ab}\,,
  \\
  &\mathcal N_{ab}=\overline{F}_{ab}+2\op i \, \frac{
{\rm Im}(F_{aI}) X^I \, {\rm Im}(F_{bJ}) X^J}{
           X^M \op{\rm Im}(F_{MN}) X^N}=\mathcal{R}_{ab}+i\,\mathcal{I}_{ab}  \,,
}
which follows from evaluating the $\mathcal N=2$ gauge kinetic function 
on the orientifold locus $X^a=0$, $F_a=0$ \cite{Grimm:2004uq}.
The period matrix in \eqref{period_m_001} is written using second derivatives of the prepotential $F$, that is
\eq{ 
\label{derivatives_001}
F_{\mathsf{IJ}} = \frac{\partial^2 F}{\partial X^{\mathsf I}\partial X^{\mathsf J}} \biggr\rvert_{X^c=F_c=0} \,,
\hspace{50pt}
\mathsf I,\mathsf J = 0, \ldots, h^{2,1}\,.
}


\subsubsection*{Some expressions}

We finally collect some technical results. 
We first define the inverse of the real part of the gauge kinetic function 
as the $\mathcal N=2$ expression restricted to the orientifold locus,
and we obtain using \eqref{derivatives_001} that
\eq{
 \bigl[(\mbox{Re}\op f)^{-1} \bigr]^{ab} = - \bigl[ (\mbox{Im}\op F)^{-1} \bigr]^{ab}\,.
}
Second, for the $\mathcal N=2$ setting 
one usually defines a $(2h^{2,1}+2)\times(2h^{2,1}+2)$ dimensional matrix $\mathcal M$
which corresponds to a metric for the third cohomology. We are interested
in the orientifold-even part evaluated on the orientifold locus 
which takes the form
\eq{
\label{matrix_m}
 \mathcal M = \left( \begin{array}{cc}
-\bigl[ ({\rm Im}\op\mathcal N)^{-1}\bigr]^{ab}  
&
 \bigl[ ({\rm Im}\op\mathcal N)^{-1}( {\rm Re}\op\mathcal N )\bigr]^a_{\hspace{4pt}b}
\\[4pt]
 \bigl[ ( {\rm Re}\op\mathcal N )({\rm Im}\op\mathcal N)^{-1}\bigr]_a^{\hspace{4pt}b}
&
 -\bigl[(  {\rm Im}\op\mathcal N) +( {\rm Re}\op\mathcal N )({\rm Im}\op\mathcal N)^{-1}( {\rm Re}\op\mathcal N)\bigr]_{ab}  
  \end{array}
  \right),
}
with $a,b=1,\ldots, h^{2,1}_+$. Note that this matrix is positive definite.  
The separate blocks of \eqref{matrix_m} can be determined in terms of 
\eqref{derivatives_001} and take the following form
\eq{
 \bigl[ ({\rm Im}\op\mathcal N)^{-1}\bigr]^{ab}  & = - \bigl[ ({\rm Im}\op F)^{-1}\bigr]^{ab} \,,
 \\
 \bigl[ ( {\rm Re}\op\mathcal N )({\rm Im}\op\mathcal N)^{-1}\bigr]_a^{\hspace{4pt}b}  & = - \bigl[ ({\rm Re}\op F)({\rm Im}\op F)^{-1}\bigr]_a^{\hspace{4pt}b} \,,
 \\
 \hspace{-5pt} 
 \bigl[(  {\rm Im}\op\mathcal N) +( {\rm Re}\op\mathcal N )({\rm Im}\op\mathcal N)^{-1}( {\rm Re}\op\mathcal N)\bigr]_{ab}  
 & = -  \bigl[ ( {\rm Im}\op F) +( {\rm Re}\op F)({\rm Im}\op F)^{-1}( {\rm Re}\op F)\bigr]_{ab}  \,.
 \hspace{-20pt}
}


\section{D3-branes wrapping three-cycles}
\label{sec_d3}

The main objects of interest for our analysis
are D3-branes wrapping three-cycles 
in the compact geometry, which  give
rise to particles in the four-dimensional 
theory.


\subsection{D-branes}

For our purpose we are looking for  objects which are charged under the $U(1)$ gauge fields $U^a$. 
Since the $U^a$ are contained in the expansion \eqref{expansion_001} of the R-R four-form $C_4$, 
we are focussing on D-branes. 


\subsubsection*{D3-branes}
\label{d3bra}

We first want to show that for the setting introduced above, the only objects coupling directly to 
the closed-string gauge fields are D3-branes wrapping three-cycles in the 
compact space.\footnote{As discussed for instance in \cite{Jockers:2005pn,Marchesano:2014bia}, the closed-string $U(1)$ gauge fields can also couple to open-string moduli. This situation is however not generic and will therefore be ignored in the following.}
To do so, we  recall the Chern-Simons part of the D-brane action. 
With $\mu_p$ its charge, $\Gamma$ the submanifold wrapped by the D-brane, 
$\mathcal F$ the open-string gauge invariant 
field strength and the $\hat{\mathcal{A}}$-genus of the curvature two-form, we have
\eq{
  \mathcal{S}_{{\rm D}p} &\supset -\mu_p \int_{\Gamma} 
    \mbox{ch}\left( \mathcal F\right) \wedge
    \sqrt{ \frac{\hat{\mathcal A}(\mathcal R_T)}{\hat{\mathcal A}(\mathcal R_N)}} \wedge
    \bigoplus_q C_q \,.
}
Since $ \mbox{ch}\left( \mathcal F\right)$ as well as the $\hat{\mathcal{A}}$-terms are even forms,
and since on a Calabi-Yau manifold the first and fifth cohomology are trivial, the only 
possible D-branes which couple to $C^{(1,3)}$ (cf.~the expansion in \eqref{expansion_001}) 
have to wrap  (orientifold-even)  three-cycles in $\mathcal X$. 
Furthermore, since $\mathcal F$ is odd under $ \Omega_{\rm P}\, (-1)^{F_{\rm L}}$
the four-dimensional part does not contain $\mathcal F$ and we arrive 
at a D3-brane wrapping a three-cycle in the compact space $\mathcal X$. 
That means, we consider D3-branes extending along the time direction in 
the non-compact four-dimensional space $\mathbb R^{3,1}$ and wrapping
three-cycles $\Gamma_3\subset\mathcal X$, where the latter can be expanded
in the basis \eqref{basis_02} as follows
\eq{
  \Gamma_3 = m_I A^I + m_a A^a + n^I B_I + n^a B_a\,.
}


\subsubsection*{Supersymmetry}

The orientifold three- and seven-planes typically present in the 
background preserve a particular combination of 
the $\mathcal N=2$ supercharges in the four-dimensional theory,
leading to $\mathcal N=1$.
However, D3-branes wrapping three-cycles in the compact space
 break supersymmetry further.
Using  the $\kappa$-symmetry formalism for D-branes 
and following the 
analysis of \cite{Bergshoeff:1996tu,Bergshoeff:1997kr,Marino:1999af,Kachru:1999vj}, 
we find the following:
\begin{itemize}
\label{page_susy}
\item A D3-brane wrapping an orientifold-even three-cycle $\Gamma_3\in H_{3+}(\mathcal X)$ breaks 
supersymmetry completely. 
The volumes of such three-cycles are expected to vanish, 
in agreement with the classification of 
toroidal orientifolds in
\cite{Lust:2006zh} where the $h^{2,1}_+$ sector belongs to the twisted sector. Alternatively, 
this conclusion is reached by projecting 
the $\mathcal N=2$ expression to 
$\mathcal N=1$.

\item A D3-brane wrapping an orientifold-odd three-cycle $\Gamma_3\in H_{3-}(\mathcal X)$ preserves
at most one-half of the $\mathcal N=1$ supersymmetry in four dimensions, that is,
such a D3-brane preserves at most two real supercharges. 
To do so, a calibration condition of the following form has to be satisfied
\eq{
  \label{calibration_001}
  \mbox{Re}\bigl(e^{i\op\theta} \Omega\op\bigr)\bigr\rvert_{\Gamma_3} = e^{-K_{\rm cs}/2}\,d\mbox{vol}(\Gamma_3) \,,
  \hspace{40pt}
  \mbox{Im}\bigl(e^{i\op\theta} \Omega\op\bigr)\bigr\rvert_{\Gamma_3}   =0\,,
}
where $\theta\in\mathbb R$ is an arbitrary phase. It has to be equal for all
such D3-branes present in the background.\footnote{
For space-time filling D-branes the angle $\theta$ is typically 
fixed 
by comparing with corresponding orientifold planes, but
such a mechanism is not available here.}

\end{itemize}
The important point is that D3-branes wrapping three-cycles in 
the Calabi-Yau orientifold do not preserve the same supersymmetry
as the orientifold three- and seven-planes. In particular, they are 
not BPS and hence the usual stability arguments do not apply ---
in general such states are therefore unstable. We come back to this
point in section~\ref{sec_stability} below.

Our findings above are consistent with the projection 
of the $\mathcal N=2$ BPS condition 
for Calabi-Yau compactifications.
More concretely, D3-branes wrapping three-cycles in Calabi-Yau 
three-folds are known to satisfy the inequality \cite{Becker:1995kb} 
\eq{
\text{Vol}(\Gamma_{3})\geq e^{K_{\rm cs}/2} 
\left| \int_{\Gamma_{3}}\Omega\op\right|  ,
\label{eq:ineqBBS}
}
where the equality corresponds to BPS states in the 
$\mathcal N=2$ theory.
When projecting this condition to $\mathcal N=1$
via the orientifold 
projection, we expect  corrections to the 
relation \eqref{eq:ineqBBS}.
However, our expectancy is that these are under control in the large complex-structure regime.
Furthermore, since these D3-branes wrap topologically non-trivial 
cycles in the Calabi-Yau orientifold while being non-BPS, 
the backreaction on the geometry should be taken into account.


\subsubsection*{Four-dimensional particles}

D3-branes wrapping three-cycles in the compact space can be interpreted as particles in the 
four-dimensional non-compact theory. 
The DBI-part of a D-brane action with $\mathcal F=0$ has the form
$\mathcal S_{{\rm D}p}\supset - T_p \int_{\Gamma} e^{-\phi}\op d\mbox{vol}$, where $T_p$ denotes
the tension of the brane. Going then to Einstein frame 
we find the following action
for D3-branes wrapping $\Gamma_3$
\eq{
  \label{action_redux_001}
  \mathcal S_{{\rm D}3} &= - T_3 \int_{\mathbb R \times \Gamma_3} 
  ds\wedge \text{dvol}(\Gamma_3) 
  - \mu_3 \int_{\mathbb R \times \Gamma_3}  C_4 
  \\[4pt]
  &\sim - m \int_{\mathbb R} 
  ds
    -  \int_{\mathbb R} \bigl(m_a \op U^a + n^a\op V_a\bigr) \,,
}
where  $ds$ is the line element of the world-line in four dimensions. We also included electric as well as magnetic couplings of the D3-brane to the gauge fields and we combine 
the corresponding charges into the vector $\mathfrak q=(m_a,n^a)$.
Depending on whether the D3-brane wraps an 
orientifold-even or an orientifold-odd
three-cycle, we find the following two classes of particles:
\begin{itemize}

\item D3-branes wrapping odd three-cycles $\Gamma_3 \in H_{3-}(\mathcal X)$ give rise
to massive particles in four dimensions. These are not charged under the $U(1)$ gauge symmetries and therefore are in general unstable. 
Using \eqref{action_redux_001} and the equality in \eqref{eq:ineqBBS} we can then read-off the mass
(in units $\sqrt{8\pi} M_{\rm Pl}$) and charges of these particles as
\eq{
\label{d3_odd}
  m = e^{K_{\rm cs}/2}
   \Bigl|m_IX^I - n^IF_I \Bigr|  + \ldots
  \,,
  \hspace{40pt} \mathfrak q = 0
  \,,
 }
  where $\ldots$ correspond to the aforementioned corrections in the orientifold setting which we expect are subleading in the large complex-structure limit.

\label{page_masses_d3}

\item D3-branes wrapping even three-cycles $\Gamma_3 \in H_{3+}(\mathcal X)$  give rise to massless four-dimensional particles.
This can be explained by noting that explicit constructions of orientifold-even three-cycles in type IIB orientifolds belong to the twisted sector and have vanishing volume \cite{Lust:2006zh}, or alternatively by projecting the $\mathcal N=2$ result.
However, these massless particles 
are charged under the $U(1)$ gauge symmetries 
with electric charge $m_a$ and magnetic charge 
$n^a$, that is
\eq{
  \label{d3_even}
  m = 0 \,, \hspace{40pt} 
  \mathfrak q = \binom{m_a}{n^a}\,.
}

\end{itemize}
D3-branes wrapping a general three-cycle can of course be massive as well as charged, 
but in these cases the $U(1)$ charges and their masses are not related to each other. 


\subsection{Stability}
\label{sec_stability}

Above we have argued
that D3-branes wrapping 
three-cycles in the compact space 
do not preserve the same supersymmetry as
the orientifold three- and seven-planes. 
They are therefore not
BPS states of the four-dimensional theory 
and thus in general unstable. 


\subsubsection*{D3-branes wrapping orientifold-odd three-cycles}

Let us start by briefly recalling the $\mathcal N=2$ situation. In this case, D3-branes wrapping special-Lagrangian three-cycles in the 
compact space
give rise to BPS states of the 
four-dimensional theory. 
Such states feel an equilibrium 
of repulsive (gauge interactions) and 
attractive (gravitational and 
scalar interactions) forces,
which allows them to be stable.
Furthermore, there can be 
infinite towers of BPS states as they do not feel a force among themselves. Nevertheless, in order to avoid decay by marginal stability the authors of  \cite{Grimm:2018ohb} use a monodromy-orbit formalism to find appropriate bound states building these towers.

In the $\mathcal N=1$ orientifold setting,
as discussed above, the gauge vectors associated to
orientifold-odd three-cycles $\Gamma_3 \in H_{3-}(\mathcal X)$ are projected out.
The D3-branes wrapping such cycles
are massive but uncharged,
and therefore feel an  attractive potential which usually manifests 
itself via a tachyonic mode and subsequent tachyon condensation and decay \cite{Sen:1999mg,Sen:1999nx}. This is in resemblance to the D-brane of the bosonic string \cite{Sen:1999mg,Sen:1999nx}, but in our case the spatial extension of the brane is wrapping three-cycles of the Calabi-Yau manifold. As a consequence,
without a proper analysis taking into account the backreaction onto the geometry
the final state is unknown.
The same reasoning applies to towers of 
such states which are again unstable, reflecting the fact that the monodromy-orbit formalism of \cite{Grimm:2018ohb} is not applicable to these uncharged D3-particles.
To summarize, D3-branes wrapping orientifold-odd three-cycles of the Calabi-Yau are in general unstable.
We make the following remarks:
\begin{itemize}

\item Around equation~\eqref{calibration_001} 
we discussed that
some D3-brane configurations preserve $\mathcal{N}=1/2$ supersymmetry in four dimensions. These states might have some notion of stability, but a 
proper analysis is beyond the scope of this work.

\item We are not considering non-BPS stability by means of K-theoretical discrete symmetries, since
states which are K-theoretical stable are not candidates to build towers of states. 
In particular, the open-string tachyon for stable non-BPS branes on orientifolds is projected out by world-sheet parity which can only happen for a single brane, whereas  towers of more than one brane are unstable and decay \cite{Sen:1999mg}.

\end{itemize}

As we have argued,  at a generic point in complex-structure moduli space
D3-branes wrapping cycles $\Gamma_3 \in H_{3-}(\mathcal X)$ are unstable. 
However, when approaching the large complex-structure limit their mass 
and their scalar interactions
approach zero and these states become 
``asymptotically stable''. 
In particular, the imbalance of forces gets reduced and  backreaction effects become less and less relevant. 
A motivation for this interpretation is the analysis of emergence for the orientifold-odd sector in section~\ref{emergencesect}.


\subsubsection*{D3-branes wrapping orientifold-even three-cycles}

D3-branes wrapping orientifold-even three-cycles $\Gamma_3 \in H_{3+}(\mathcal X)$ in the compact space
correspond to massless-charged particles in 
the four-dimensional theory, which 
in the context of the swampland program 
have been discussed for instance in \cite{Heidenreich:2017sim,Heidenreich:2019zkl}.
Many of these particles are however unstable and 
can decay into their elementary constituents, since
there is no mass associated to them. The stable states correspond to the following fundamental charges (for a fixed $a$):
\eq{
\label{states_elem}
\arraycolsep2pt
\begin{array}{lcl@{\hspace{30pt}}l}
(m_a,n^a)&=&(1,0) & \mbox{massless electric particles,}
\\[4pt]
(m_a,n^a)&=&(0,1) & \mbox{massless  magnetic monopoles.}
\end{array}
}
This reasoning is supported from the $\mathcal{N}=2$ perspective where it was argued in \cite{Strominger:1995cz} that the conifold singularity is resolved by the addition of the electric state $(m_a,n^a)=(1,0)$, while the states $(m_a,n^a)=(\mathbb Z,0)$ should decay to the previous one in order to match the one-loop beta function for the electric gauge coupling. 
Since the orientifold locus $X^a=0$ resembles a
conifold singularity, 
it is not surprising that we inherit these  stable states in the orientifold theory.


\subsubsection*{Remark}

We close this section be remarking that D3-branes wrapping general three-cycles can of course be massive as well as charged, 
but in these cases the $U(1)$ charges and their masses are not related to each other. As a consequence, they ultimately decay to the previous cases.


\section{Swampland conjectures}
\label{sect_swampconj}

We now want to test several swampland conjectures
for particles in the setting introduced above. 
These are the weak gravity conjecture, 
the swampland distance conjecture and the emergence 
proposal.


\subsection{Weak gravity conjecture}

We begin our discussion of swampland conjectures
with the weak gravity conjecture. 


\subsubsection*{Recalling the conjectures}

We start by briefly recalling the electric weak gravity conjecture \cite{ArkaniHamed:2006dz} in its modern formulation \cite{Palti:2019pca}:
\begin{itemize}

\item[] Consider a $U(1)$ gauge theory  with gauge coupling $g$ 
coupled to gravity and described by the following action
\begin{equation}
\mathcal S= \frac{1}{2} \int_{\mathbb R^{3,1}} \left[
M_{\rm P}^2 \, R\star 1 - \frac{1}{2\op g^2}\op 
F\wedge\star F + \ldots\op \right],
\label{eq:actionWGC}
\end{equation}
where $R$ denotes the Ricci scalar and $F$ is the $U(1)$ field-strength two-form.
Then there exists a particle in the theory with mass $m$ and charge $\mathfrak q$ satisfying the inequality 
\begin{equation}
m
\leq \sqrt{2}\op \mathfrak q\op g
M_{\rm P} \, .
\label{eq:eWGC}
\end{equation}

\end{itemize}
This conjecture has been refined in many ways, and for our purpose the weak gravity conjecture with scalar fields will be relevant \cite{Palti:2017elp}.
We recall this conjecture as follows:
\begin{itemize}

\item[] Consider a gravity theory with massless scalar fields $z^i$ and $U(1)$ gauge fields $U^a$ described by the
action \eqref{action_001}. Then there should exist a particle with mass $m(z)$ satisfying the bound
\begin{equation}
\label{conjecture_wg}
\mathcal{Q}^2
\geq m^2+G^{i\ov j}(\partial_{z^{i}} m)(\partial_{\ov z^{j}}m) \,, \hspace{40pt}
\mathcal Q^2 = \frac{1}{2} \op\mathcal Q^T \mathcal M \op\mathcal Q\,,
\end{equation}
where $\mathcal Q$ denotes the vector 
$\mathcal Q = (m_I,m_a,n^I,n^a)$,
$\mathcal M$ is the positive-definite matrix similar to the one 
defined in \eqref{matrix_m} and
$G^{i\ov j}$ is the inverse of the K\"ahler metric 
\eqref{metric_k}.

\end{itemize}


\subsubsection*{Verifying the conjectures}

We now discuss how these conjectures are satisfied in the orientifold setting introduced in the previous section:
\begin{itemize}

\item For D3-branes wrapping odd three-cycles $\Gamma_3 \in H_{3-}(\mathcal X)$ 
we recall from  \eqref{d3_odd} that their $U(1)$ charges 
$\mathfrak q=(m_a,n^a)$
are vanishing. 
These particles do not provide the required charged particle
for the weak gravity conjecture \eqref{eq:eWGC}, but
one can check that they 
verify the scalar weak gravity conjecture  \eqref{conjecture_wg}
without $U(1)$ gauge interactions \cite{Palti:2017elp}.

\item D3-branes wrapping even three-cycles $\Gamma_3 \in H_{3+}(\mathcal X)$ 
are massless and their charges $\mathfrak q$ are given by the wrapping numbers
$(m_a,n^a)$. 
The electric weak gravity conjecture \eqref{eq:eWGC} is therefore verified. Furthermore, using \eqref{d3_even} we find that
\eq{
  \frac{1}{2}\op \binom{m}{n}^T \mathcal M\op \binom{m}{n} \geq 0 \,,
}
which is satisfied since the matrix $\mathcal M$ is positive definite.

\end{itemize}
We therefore conclude that the electric weak gravity conjecture
\eqref{eq:actionWGC}
and  weak gravity conjecture with scalar fields \eqref{conjecture_wg}
are trivially satisfied.


\subsubsection*{Completeness conjecture and tower versions}

The statement of the completeness conjecture is that a gravity theory with a gauge symmetry must have states of all possible charges under that gauge symmetry \cite{Polchinski:2003bq}. 
In our case this conjecture is satisfied by D3-branes wrapping orientifold-even cycles $\Gamma_3\in H_{3+}(\mathcal X)$ with wrapping numbers $(m_a,n^a)$.
However, in general these states are not stable against decay to the elementary ones \eqref{states_elem}
and hence the completeness conjecture is only satisfied
as long as stability is not required. 
The completeness conjecture  motivates the tower versions of the weak gravity conjecture \cite{Heidenreich:2015nta,Heidenreich:2016aqi,Andriolo:2018lvp}, which suggests that for the D3-brane particles discussed here the tower versions are satisfied only if stability of the states is relaxed.

Furthermore, 
in \cite{Palti:2019pca} it is argued that avoidance of generalized global symmetries for an $U(1)$ gauge theory coupled to gravity requires at least one single charged state. Remarkably, in \cite{Gaiotto:2014kfa} it is claimed that an extension of the previous argument to discrete generalized global symmetries requires all possible charges in order to break any possible discrete symmetry. It would be interesting to explore the compatibility between our results concerning the completeness conjecture and \cite{Gaiotto:2014kfa}.


\subsubsection*{Summary and remark}

We now want to briefly summarize our findings 
regarding the 
weak gravity conjecture and make the following remark:
\begin{itemize}

\item We have shown that the weak gravity conjectures 
for particles are verified 
in type IIB orientifold compactifications
with O3- and O7-planes 
for D3-branes wrapping three-cycles.
However, the tower versions  cannot be claimed to be verified as long as stability is required.

\item Analogously to type IIB orientifolds with O3-/O7-planes case studied here, the same results are obtained for type IIB orientifolds with O5-/O9-planes and for type IIA orientifolds.

\item In \cite{Font:2019cxq} it was shown that 
in the orientifold
setting
BPS D5-branes wrapping three-cycles  verify the weak gravity conjecture as well. These are however
extended objects and not particles, and they 
are charged under a different gauge symmetry. 

\end{itemize}


\subsection{Swampland distance conjecture}

We next consider the swampland distance conjecture
for type IIB orientifolds with O3-/O7-planes 
introduced above.


\subsubsection*{Recalling the conjecture}

The swampland distance conjecture 
has been formulated in \cite{Ooguri:2006in}, and
has been 
refined in \cite{Klaewer:2016kiy}. The refined version reads as follows:
\begin{itemize}

\item[] Consider a theory coupled to gravity with a moduli space $\mathcal M_{\rm mod}$ parametrized
by the expectation values of some fields without potential. Let the geodesic distance between any two 
points $P,Q \in \mathcal{M}_{\rm mod}$ be denoted $d(P, Q)$. 
If $d(P,Q) \gtrsim M_{\rm P}$, then there exists an infinite tower of states with mass scale $m$ such that
\begin{equation}
m(Q)<m(P)\,e^{-\mu\frac{d(P,Q)}{M_{\rm P}}}\,,
\label{eq:RSDC}
\end{equation}
where $\mu$ is some positive constant of order one. 
This statement holds even for fields with a potential, where the moduli space is replaced with the field space in the effective theory.

\end{itemize}
We now want to verify this conjecture for the setting of the previous section. In particular, we
are starting from a generic point in complex-structure moduli space  and approach a large complex-structure point. 
The results obtained in \cite{Grimm:2018ohb} for the $\mathcal N=2$ theory suggest that also 
in our $\mathcal N=1$ orientifold setting D3-branes wrapping three-cycles give rise to the 
infinite tower of states.
\begin{itemize}

\item We have argued that the mass of D3-branes wrapping even three-cycles $\Gamma_3 \in H_{3+}(\mathcal X)$ 
vanishes and hence they do not play a role for the swampland distance conjecture.

\item For D3-branes wrapping odd three-cycles $\Gamma_3 \in H_{3-}(\mathcal X)$ 
the $U(1)$ charges are vanishing but they are massive (cf.~equation \eqref{d3_odd}).
These states are therefore of interests for 
the swampland distance conjecture.

\end{itemize}


\subsubsection*{Geodesic distance}

Let us consider a generic point in complex-structure moduli 
space for which the one-loop and non-perturbative corrections to the prepotential 
\eqref{prepotential_full} can be ignored. 
The K\"ahler metric is computed via \eqref{metric_k}, and the geodesic equation for
such K\"ahler geometries reads in general
\eq{
  \label{geodesic_001}
  0 = \ddot z^i + \Gamma^i_{jk} \dot z^j \dot z^k\,,
  \hspace{50pt}
  \Gamma^i_{jk}  = G^{i\ov m} \partial_j G_{k\ov m} \,,
}
where a dot indicates a derivative with respect to the parametrization $t$ of the geodesic. 
The geodesic distance is defined as
\eq{
  d(P,Q) = \left\lvert\int_{t_1}^{t_2} dt \sqrt{2\op  \dot z^i\, G_{i\ov j}\, \dot{\ov{z}}{}^{\ov j}} \right\rvert
  \,,
  \hspace{50pt} 
  \arraycolsep2pt
  \begin{array}{lcl}
  z^i(t_1) &=& P \in \mathcal{M}_{\rm mod}\,,
  \\[4pt] 
  z^i(t_2) &=& Q \in \mathcal{M}_{\rm mod}\,.
  \end{array}
}
For simplicity we now restrict ourselves to a situation 
with one complex-structure modulus, that is $h^{2,1}_-=1$.
We use the notation $z^1\equiv z=u+i\op v$, and 
the moduli space is a  hyperbolic space with metric
\eq{
  G_{1\ov 1} = \frac{3}{4}\op\frac{1}{v^2} \,.
}
The solutions to the geodesic equation \eqref{geodesic_001}
are well known, and correspond to 
lines with constant $u$ and to circles intersecting $v=0$ at
a right angle
\eq{
  \label{geos_004}
  z_{(1)} = \alpha + i\op \beta \op e^{ t}
  \,,
  \hspace{50pt}
  z_{(2)} = \alpha + \frac{\beta}{\cosh(t)}\op 
  \bigl[ \sinh( t) + i \bigr]\,,
}
where $\alpha,\beta={\rm const}$ and $\beta>0$.
The geodesic distance for these two cases is computed as follows
\eq{
\label{geodist}
  d_{(1)}(P,Q) &
  = \sqrt{\frac{3}{2}}\,\lvert t_Q-t_P\rvert
  =
  \sqrt{\frac{3}{2}} \left\lvert \log \frac{v_Q}{v_P} \right\rvert
  \,,
  \\
  d_{(2)}(P,Q) &= \sqrt{\frac{3}{2}}\,\lvert t_Q-t_P\rvert\,.
}


\subsubsection*{Verifying the conjecture}

As mentioned above, motivated by the $\mathcal N=2$ 
results we expect that  D3-branes wrapping 
orientifold-odd three-cycles will provide the tower 
of
states which becomes exponentially light. 
In our example with $h^{2,1}_-=1$ we see that
D3-branes with wrapping numbers $n^0\neq0$ or $n^1\neq0$ 
do not satisfy the swampland distance conjecture \eqref{eq:RSDC}
in the large complex-structure limit $v\to\infty$;
they are expected to provide the required states 
in the small complex-structure limit.
We focus here on the large complex-structure limit and D3-branes with wrapping numbers $m_0$ and $m_1$:
\begin{itemize}

\item We first consider D3-branes satisfying the 
calibration condition \eqref{calibration_001}.
In this case the mass is given by
\eq{
  m(z)= \frac{1}{\sqrt{8\op d_{111}v^3}}\,
  \mbox{Re}\left[ e^{i\hat\theta} \bigl(
  m_0 + m_1 z + n^0 d_{111} z^3 - 3\op n^1 d_{111} z^2 \bigr) \right],
}
with $d_{111}$ the triple intersection number 
introduced in \eqref{prepotential_full} and where we defined
$\hat\theta = \theta + \arg( X^0)$.
If we approach the large complex-structure limit
via the first geodesic in \eqref{geos_004}, 
it turns out that there
are only two choices for $\hat\theta$ compatible 
with the calibration condition. The tower satisfying 
the swampland distance conjecture
together with the constant $\mu$ 
are the following
\eq{
\arraycolsep2pt
\begin{array}{@{}lcl@{\hspace{30pt}}lcl@{\hspace{15pt}}lcl@{\hspace{15pt}}lcl@{\hspace{30pt}}lcl@{}}
\hat\theta & = & 0\,, & m_0 & > & 0\,, & m_1 & = & 0\,, 
& n^a&=&0 \,, 
& \mu &=& \sqrt{3/2}\,,
\\[4pt]
\hat\theta & = & \pi/2\,, & m_0 & = & 0\,, & m_1 & > & 0\,, 
& n^a&=&0 \,, 
& \mu &=& \sqrt{1/6}\,.
\end{array}
}

\item Next, we relax the calibration condition and
consider states satisfying the equality in
\eqref{eq:ineqBBS}. 
These are the states inherited from the 
$\mathcal N=2$ theory, and their mass  is given by
\eq{
  m(z) = \frac{1}{\sqrt{8\op d_{111}v^3}}\,
  \Bigl|
  m_0 + m_1 z + n^0 d_{111}z^3 
  -3\op n^1 d_{111}z^2 \Bigr|\,.
  \label{eq:mag}
}
Approaching the large 
complex-structure limit again via
first geodesic in \eqref{geos_004}, we see
that tower of states satisfying 
the swampland distance conjecture are given by
\eq{
\arraycolsep2pt
\begin{array}{@{}lcl@{\hspace{30pt}}lcl@{\hspace{15pt}}lcl@{\hspace{15pt}}lcl@{\hspace{30pt}}lcl@{}}
\hphantom{\hat\theta} & \hphantom{=} & \hphantom{\pi/2\,,} & m_0 & \in & \mathbb Z\,, & m_1 & \in & \mathbb Z\,, 
& n^a&=&0 \,, 
& \mu &=& \sqrt{1/6}\,.
\end{array}
}

\item We also want to consider the second
geodesic in \eqref{geos_004}, for which  the 
large complex-structure limit is 
characterized by $\beta\ggg 1$ and $t\to 0$. 
For this geodesic we do not find any 
tower of D3-brane states satisfying the
swampland distance conjecture.
However, since this path is along the boundary of
the complex-structure moduli 
space it is not clear whether the
conjecture applies. It would be interesting to 
study this point further.

\end{itemize}
To summarize, along the path $u=\textrm{const.}$ and
$v\to\infty$ we have verified  the swampland distance conjecture for D3-particles in type IIB orientifold
compactifications, however, the tower of 
states is in general not stable. 
Furthermore, we were not able to verify the 
swampland distance conjecture for a path 
along the boundary of complex-structure moduli space.


\subsubsection*{Relation between swampland distance
and weak gravity conjectures}

For $\mathcal N=2$ theories originating from type IIB 
Calabi-Yau compactifications it was found 
in \cite{Grimm:2018ohb} 
that the swampland distance as well as the
weak gravity conjecture are verified by BPS D3-branes wrapping special Lagrangian three-cycles. 
However, for $\mathcal N=1$ theories obtained via 
an orientifold projection we have noted the $h^{2,1}$  vector multiplets in $\mathcal N=2$ 
are projected to $h^{2,1}_-$ 
chiral and $h^{2,1}_+$  vector multiplets in 
$\mathcal N=1$ (c.f.~equation \eqref{multiplets_split}). In particular, 
in the orientifold theory both multiplets are independent 
of each other, and there exist for instance orientifolds 
with $h^{2,1}_+=0$. 
Thus, D3-brane particles which are needed for the swampland distance conjecture are uncharged and trivially do not satisfy the electric weak gravity conjecture \eqref{eq:eWGC} --- and charged D3-brane particles satisfy the weak gravity conjecture but are not needed for the swampland distance conjecture. 
Thus, the relation between the weak gravity and
swampland distance conjectures 
(and emergence to be discussed below) 
is determined by the bulk supergravity spectrum.

This observation extends to similar cases for which D-branes couple to a bulk spectrum in $\mathcal N=2$
which splits under the orientifold projection into different $\mathcal N=1$ multiplets. For example, in K\"ahler moduli 
space the same happens for D-branes wrapping two-cycles (e.g.~the D-strings studied in \cite{Font:2019cxq}) coupling to both $J$ and $C_{2}$. It is also worth mentioning that 
similar to the complex-structure moduli, also the K\"ahler twisted sector of the examples in \cite{Lust:2006zh} contains the $h_{-}^{1,1}$ cycles, so we observe an analogy with respect to the situation studied in this work.


\subsection{Emergence proposal}
\label{emergencesect}

In this section we discuss the 
emergence proposal for D3-branes wrapping 
three-cycles in
type IIB orientifold compactifications
with orientifold three- and seven-planes.


\subsubsection*{Recalling the proposal}

The emergence proposal has been formulated in
\cite{Grimm:2018ohb,Palti:2019pca}, which 
we recall as follows:
\begin{itemize}

\item[] The dynamics (kinetic terms) for all fields are emergent in the infrared by 
integrating out towers of states down from an ultraviolet scale $\Lambda_{\rm UV}
$ which is 
below the Planck scale.

\end{itemize}
At a practical level this means that at an UV scale $\Lambda_{\rm UV}$, the renormalization group flow has a boundary condition on all  fields forcing them to have vanishing kinetic terms. 
Let us make the following remarks concerning the emergence proposal:
\begin{itemize}

\item For towers of states with equidistant mass and charge
separation and without taking care of higher-loop corrections, it has been argued in \cite{Palti:2019pca}
that imposing emergence of gravity 
one obtains 
the species scale  $\Lambda_{\rm s}$ 
\cite{Dvali:2007hz} as the UV scale.

\item Imposing emergence of a gauge field and a 
scalar field  (to one-loop order) one 
recovers the magnetic weak gravity and the swampland 
distance conjecture, respectively. 
In \cite{Grimm:2018ohb} this mechanism has been 
discussed for type IIB Calabi-Yau compactifications 
with towers of states given by D3-branes wrapping 
special Lagrangian three-cycles.

\item For orientifold compactifications with $\mathcal N=1$ 
supersymmetry in four dimensions we expect a modified picture. Indeed, the non-renormalization theorem for $\mathcal{N}=2$ changes in the case of $\mathcal{N}=1$ orientifolds as the K\"ahler potential $K_{\rm cs}$ receives loop corrections in terms of Eisenstein series 
 \cite{Haack:2017vko}. 
These corrections are however subleading and expected to be irrelevant as we approach the large complex-structure limit. Besides, the gauge kinetic functions remain one-loop exact.

\end{itemize}


\subsubsection*{Emergence for the orientifold-odd 
third homology}

We now analyze the emergence proposal for the 
orientifold-odd third cohomology.
As discussed above, D3-branes wrapping 
cycles $\Gamma_3\in H_{3-}(\mathcal X)$ are massive but unstable.
Integrating out these states gives rise to logarithmic
corrections to the complex-structure moduli-space metric
of the schematic form
\eq{
  \label{running_1}
  G_{i\ov j} \bigr\rvert_{\rm IR}
  = G_{i\ov j} \bigr\rvert_{\rm UV}
  + c_- \sum_{\alpha} \bigl(\partial_{z^i} m^{(\alpha)}\bigr)
  \bigl(\partial_{\ov z^j} m^{(\alpha)}\bigr) \log\frac{\Lambda_{\rm UV}}{m^{(\alpha)}} \,,
}
where $c_-$ is a normalization constant and the sum runs over all 
D3-brane states with mass below the cutoff scale $\Lambda_{\rm UV}$.
Following the argumentation of \cite{Grimm:2018ohb,Palti:2019pca}
for the emergence proposal 
in the $\mathcal N=2$ setting, 
the metric $G_{i\ov j}\rvert_{\rm UV}$ in the UV vanishes while 
the sum over logarithmic corrections generates the 
metric $G_{i\ov j}\rvert_{\rm IR}$ in the IR.
It is beyond the scope of this work to check \eqref{running_1} 
explicitly for our setting, but we would like to 
make the following comments:
\begin{itemize}

\item For the orientifold setting studied in this work 
we have argued that the tower of D3-brane states is in general unstable. In order to verify the emergence proposal 
we therefore have to integrate-out (infinite) unstable states, 
which are however expected to become asymptotically-stable in the 
large complex-structure limit.

\item In \cite{Font:2019cxq}
it has been proposed that also four-dimensional domain 
walls contribute to the running of the moduli-space metric. These domain walls are given by 
BPS D5-branes wrapping three-cycles 
in the compact space, but to our knowledge it is not known how to integrate out extended objects at the quantitative level. 
This ambiguity is partially reflected in the unspecified 
constant $c_-$ in \eqref{running_1}.

\item It is curious to note that
in the $\mathcal N=1$ setting we have in 
general all-order loop corrections to the K\"ahler potential 
but also that the D3-brane particles being integrated out are 
not stable.

\end{itemize}
We believe these questions deserve further 
investigation in the future.


\subsubsection*{Emergence for the orientifold-even 
third homology}

We now consider D3-branes wrapping orientifold-even 
three-cycles $\Gamma_3\in H_{3+}(\mathcal X)$ 
in the compact space. These states are
massless but are charged under the closed-string 
$U(1)$ gauge fields $U^a$,
and therefore contribute to one-loop 
corrections to the gauge kinetic function.
In particular, in field theory we have
\eq{
  \label{running_2}
  \mbox{Re}\op f^{(\rm f)}_{ab}\op \bigr\rvert_{\rm IR}
  &= 
  \mbox{Re}\op f^{(\rm f)}_{ab}\op \bigr\rvert_{\rm UV}
  + c_+ \lim_{m^{(\alpha)}\to0}
  \sum_{\alpha} \mathfrak q^{(\alpha)}_a
  \mathfrak q^{(\alpha)}_b \log\frac{\Lambda_{\rm UV}}{m^{(\alpha)}} \,,
}
where the sum is over all D3-brane particles
labelled by $\alpha$,
$\mathfrak q^{(\alpha)}_a$ denote their electric 
charges and $m^{(\alpha)}$ denote their masses which vanish at the orientifold locus.
The constant $c_+$ is a normalization constant, 
and the superscript indicates 
that this is a field-theory result.
As expected,
we obtain logarithmic divergences  due to massless particles running in the loop.\footnote{There is an ambiguity in performing 
the sum over all D3-brane particles with mass below $\Lambda_{\rm UV}$ and taking the 
limit $m^{(\alpha)}\to 0$ 
in \eqref{running_2}. When first summing 
and then taking the limit, one 
obtains a polynomial divergence.}

Let us now turn  to the string-theory computation. 
Using the relation \eqref{period_m_001} and the explicit form 
of the prepotential \eqref{prepotential_full},
we find at the orientifold locus $X^a=0$ (without imposing $F_a=0$) 
\eq{
  \label{lukas}
  \mbox{Re}\op f^{({\rm s})}_{ab} = \mbox{Im}\op F_{ab}
  = 6\op d_{abi} v^i + (2\pi)^2 \sum_k k_a k_b \op n_k \op
  \mbox{Re} \log\left( 1 - e^{2\pi i k_i z^i}
  \right).
}
Here we used again $v^i = \mbox{Im}\op z^i$
and the superscript indicates that this is  
the (effective) string-theory result.
Following now the philosophy of the emergence proposal \cite{Grimm:2018ohb,Palti:2019pca},
the gauge-kinetic function should vanish in 
the ultraviolet and the expression in 
the infrared is generated by loop corrections. 
In the present situation we argue as follows:
\begin{itemize}

\item The gauge-kinetic function in the infrared 
is given by the expression \eqref{lukas}.
Let us then recall our 
discussion from 
page~\pageref{page_or_sol}
and consider the solution 
\eqref{or_sol_2} to the orientifold condition $F_a=0$. Reinstalling the $X^a$-dependence
we have
\eq{
  \label{lll}
  \mbox{Re}\op f^{({\rm s})}_{ab} 
  = 6\op d_{abi} v^i + (2\pi)^2 
  \lim_{X^a\to 0}\sum_k k_a k_b \op n_k \op 
  \mbox{Re} \log\left( - 2\pi i\op k_a X^a/X^0 \right) .
}
Recalling furthermore our results for the masses from page~\pageref{page_masses_d3} and performing a similar analysis for 
D3-branes wrapping orientifold-even cycles, we find for electri\-cally-charged particles that
$m=\lim_{X^a\to 0}e^{K_{\rm cs}/2} | m_a X^a |$
(in units of $M_{\rm Pl}$) with $m_{a}$ given by \eqref{states_elem}. 
Using this expression in \eqref{lll}, 
we obtain up to prefactors and regular terms 
\eq{
  \label{das_boot}
  \mbox{Re}\op f_{ab}^{({\rm s})} 
  \sim 
  \lim_{m^{(\alpha)}\to0}
  \sum_{\alpha} k^{(\alpha)}_a 
  k^{(\alpha)}_b 
  \log \frac{M_{\rm s}}{m^{(\alpha)}}\,,
}
where $M_{\rm s}$ denotes the string scale
and where we relabeled the vectors $k$ appearing 
in the sum by $\alpha$.
This expression has the same form
as the field-theoretical one-loop correction
shown in \eqref{running_2}, and thus 
it is feasible that the emergence proposal is 
verified.

\item The above-mentioned string-theory result
for the gauge-kinetic function is not complete.
Indeed, the results of \cite{Strominger:1995cz}
imply that \eqref{lukas} receives corrections in 
the ultraviolet from D3-branes wrapping once around orientifold-even cycles.
Properly taking into account these states
cure the logarithmic divergence
in the Calabi-Yau moduli space, and hence
the gauge-kinetic functions
\eqref{lukas} and \eqref{lll} are expected to be 
finite in a quantum-gravity theory.
In particular, in the infrared we expect 
from string theory a behavior 
of the form
\eq{
\label{ffull}
\mbox{Re}\op f^{({\rm full})}_{ab}\op \bigr\rvert_{\rm IR}
= 6\op d_{abi} v^i + \mbox{finite}\,,
}
with the superscript indicating the full expression.

\end{itemize}
To summarize, we have argued that the 
emergence proposal for D3-branes wrapping orientifold-even three-cycles is feasible to 
be verified, 
though we have presented here only qualitative analysis and a more thorough check including numerical factors and regular terms would be necessary.


\subsubsection*{Remarks and open questions}

Let us make the following remarks 
concerning our discussion of 
emergence in the orientifold-even sector:
\begin{itemize}

\item In our analysis we have encountered
a difference between the field-theory and 
the string-theory computation.
In particular, the field-theory 
expression  \eqref{running_2} for 
the gauge-kinetic function is logarithmically
divergent whereas the full string-theory 
expression \eqref{ffull} 
is expected to be finite.

\item The gauge-kinetic function
\eqref{ffull} is the expression 
expected from string theory 
once the effect of D3-branes wrapping orientifold-even cycles on the moduli-space
geometry has been taken into account. 
For the emergence proposal this gauge coupling
in the infrared should be obtained entirely 
from loop-corrections with charged particles
running in the loop, however,
in our case these charged particles are massless and do not couple to the complex-structure moduli.
It is therefore not clear to us how the $v^i$
as well as $d_{abi}$ dependence in \eqref{ffull} can be obtained.
\begin{itemize}

\item A potential answer to this question is that there exist particles which couple simultaneously to the $U(1)$ gauge fields and to the complex-structure moduli, and where the coupling depends on $d_{abi}$, such that generate the linearly divergent term when taking into account their effect in \eqref{running_2}. However, there are no other known states with this property.

\item Another possibility is that the triple intersection numbers $d_{abi}$ vanish. This appears to be a too-strong requirement, since in T-dual settings these can be non-vanishing. 

\item Since D3-branes
wrapping orientifold-even three-cycles 
do not preserve the bulk supersymmetry (see
our discussion on page~\pageref{page_susy}),
the gauge-kinetic function may 
receive higher-loop corrections which
can generate a term linear in $v^i$.

\end{itemize}

\item We also note that when taking the large complex-structure limit
in \eqref{ffull}, the gauge symmetry becomes global. 
Although there exists an obstruction against reaching the limit $v^i\to \infty$ due to 
towers of orientifold-odd D3-branes, it is unexpected to find a global gauge 
symmetry associated with integrating out states
uncharged under it. 
In particular, this behavior is usually associated with integrating out a tower of states which becomes massless as in \eqref{eq:RSDC} and which is charged under the gauge symmetry. However, in our setting we do not realize a possible tower of states verifying both properties.

\end{itemize}
We leave open these questions here 
and refer for a more detailed description 
and further instructive attempts to solve them
to section~6 of \cite{martin:2019}.


\section{Summary and conclusions}
\label{sect_conc}

In this paper we have studied swampland
conjectures for type IIB orientifolds with 
O3- and O7-planes. We have allowed
for orientifold projections with 
$h^{2,1}_+\neq 0$, which leads to 
closed-string $U(1)$ gauge fields in 
four dimensions. 
The weak gravity conjecture, the swampland
distance conjecture and emergence proposal 
have been investigated in the context 
of the $\mathcal N=2$ parent theory
in \cite{Grimm:2018ohb}, and
here we were interested in the question 
\textit{What happens to the $\mathcal N=2$ 
analysis when we perform 
an orientifold projection?}


\subsubsection*{Summary}

Let us summarize the main results 
of our analysis:
\begin{itemize}

\item The relevant objects for the
above-mentioned swampland 
conjectures in our setting are D3-branes
wrapping three-cycles in the compact space.
In the $\mathcal N=2$ theory these 
couple to vector multiplets, and they give rise 
to four-dimensional particles charged under
$U(1)$ gauge fields and with mass depending
on the complex-structure moduli.
In the orientifold theory the 
$\mathcal N=2$ vector multiplets split into
vector and chiral multiplets of $\mathcal N=1$
supergravity 
as illustrated in \eqref{multiplets_split}.
Correspondingly, D3-branes can be separated
into massive-uncharged (orientifold-odd) and massless-charged (orientifold-even)
particles in four dimensions. 
While for $\mathcal N=2$ these states can be 
BPS and therefore can be stable, for the $\mathcal N=1$ 
theory they are in general unstable
and non-supersymmetric. This suggests 
that the $\mathcal N=1$ swampland conjectures 
studied in this paper are satisfied only for 
in general 
unstable states.

\item We have illustrated that 
the weak gravity conjecture
\cite{ArkaniHamed:2006dz}  and the weak gravity 
conjecture with scalar fields
\cite{Palti:2017elp} are trivially satisfied 
in the $\mathcal N=1$ setting.
This is true both for D3-branes wrapping orientifold-odd and 
orientifold-even three-cycles, 
however, the tower versions can only be satisfied
if the stability requirement is relaxed.

\item We have verified the swampland distance
conjecture \cite{Ooguri:2006in,Klaewer:2016kiy}
for D3-branes wrapping orientifold-odd
three-cycles in the Calabi-Yau three-fold. 
For the case $h^{2,1}_-=1$ we have studied two 
types of geodesics, where one geodesic 
along the boundary of the complex-structure moduli 
space appears to violate the conjecture. 
We noted furthermore that 
the massless-charged D3-brane particles do not 
contribute to the swampland distance conjecture,
and that the tower of states
is only asymptotically stable.

\item For the emergence proposal \cite{Grimm:2018ohb,Palti:2019pca}
we have seen that integrating out 
D3-branes wrapping 
orientifold-odd three-cycles gives
rise to the correct behavior of the 
K\"ahler metric for the complex-structure moduli.
The analysis in the orientifold-even sector 
was more ambiguous, but we argued that 
D3-branes wrapping (collapsing) three-cycles 
can re-produce the leading divergency 
in the gauge-kinetic function for the closed-string 
$U(1)$ gauge fields. However, 
when properly taking into account 
the effect of such D3-branes on the moduli-space
geometry it is expected that this divergence
is removed.
There is furthermore an open question of how 
the leading regular term in
the gauge-kinetic function 
can be obtained from loop corrections, 
and we briefly discussed possible solutions.

\end{itemize}


\subsubsection*{Open questions and future directions}

We  comment on open questions and future 
research directions:
\begin{itemize}

\item The closed-string 
gauge theories originating from 
the orientifold-even sector 
have been studied mostly from 
a supergravity point of view
and require a better understanding
within string theory.
More concretely, the implications 
of the orientifold locus 
$X^a=0$, $F_a=0$ on the world-sheet 
instanton corrections 
(c.f.~page~\pageref{page_or_loc}) need
to be clarified, 
the properties of D3-branes wrapping 
three-cycles $\Gamma_3 \in H_{3+}(\mathcal X)$
deserved to be  studied further, and
the emergence proposal 
for the gauge theories should be made 
more explicit. 
To do so a concrete example would be helpful,
for which the results in \cite{Lust:2006zh} 
provide
a good starting point.

\item While in the $\mathcal N=2$ setting 
D3-branes wrapping three-cycles in 
the compact space can be BPS, this property
is lost after orientifolding. More concretely, the corresponding states 
in the four-dimensional theory no longer
preserve supersymmetry
and we expect corrections to
the mass formula for orientifold-odd 
D3-branes shown in \eqref{d3_odd},
associated with their instability and with 
backreaction effects onto the geometry.
We have argued that these effects should be subleading as we approach the infinite-distance limit and have disregarded them, however, such corrections should be  computed explicitly.

This is in line with results obtained in \cite{Blumenhagen:2019vgj}, where logarithmic 
quantum
corrections to various swampland conjectures -- including the de-Sitter swampland conjecture -- have been discussed. 
At the same time, the authors of  \cite{Ooguri:2018wrx} have connected the 
de-Sitter conjecture to the  swampland distance and weak gravity conjectures. 
If these connections survive in the orientifold setting, it would be interesting to see if through them 
the work of \cite{Blumenhagen:2019vgj} 
can shine light on the concrete form of possible modifications to the mass formula \eqref{d3_odd}, to the swampland distance and to  weak gravity conjectures for our setting.

\item In this work we focused on type IIB orientifolds with O3- and O7-planes. 
It would be interesting to extend our discussion to type IIA orientifolds where a similar splitting of the $\mathcal N=2$ vector multiplet occurs \cite{Grimm:2004ua}. From mirror symmetry we expect that the mirror towers/D-branes satisfy a similar behavior, which for the $\mathcal{N}=2$ case was shown in \cite{Corvilain:2018lgw}. Besides, the E2-instantons analyzed in \cite{vittmann:2019} have properties similar to the case we studied but in the complex-structure moduli space of type IIA.
Furthermore, in the work of \cite{Font:2019cxq} it would be interesting to explore the role of the uncharged towers of extended objects.

\item Many studies of the swampland 
conjectures are based on $\mathcal{N}\geq2$
supergravity theories (see \cite{DallAgata:2020ino} 
for 
recent work with extended supersymmetry).
However, in order to make contact with 
phenomenology, theories with $\mathcal{N}=1$
supersymmetry are more suitable. 
What we find in our work is that
many of the connections between the swampland conjectures and the emergence proposal are
modified for $\mathcal{N}=1$, and therefore
further studies of these connections 
and their implications for phenomenology
would be desirable. 

\end{itemize}


\vskip2em
\subsubsection*{Acknowledgments}

We would like to thank 
R.~Blumenhagen, 
L.~Martucci and 
E.~Palti for very helpful discussions
and we thank D.~L\"ust for support.
EP thanks 
K.~Schalm 
for hospitality
and
L.~Plauschinn
for enlightening discussions. 
MER is grateful to H.~Erbin for suggestions on literature concerning tachyon condensation in string field theory \cite{Sen:1999nx} and non-BPS stability \cite{Sen:1999mg} as well as to H.~\'Asmundsson, D.~Bockisch, P.~Fragkos, D.P.~Lichtig, A.~Makridou, I.~Mayer, S.P.~Mazloumi and J.D.~Sim\~ao for patient and useful discussions and finally to his family and friends for unconditional support along the period of elaboration of this work.


\clearpage
\nocite{*}
\bibliography{references}

\providecommand{\href}[2]{#2}\begingroup\raggedright\begin{thebibliography}{10}

\bibitem{Vafa:2005ui}
C.~Vafa, ``{The String landscape and the swampland},''
  \href{http://xxx.lanl.gov/abs/hep-th/0509212}{{\tt hep-th/0509212}}.

\bibitem{Palti:2019pca}
E.~Palti, ``{The Swampland: Introduction and Review},'' {\em Fortsch. Phys.}
  {\bf 67} (2019), no.~6 1900037,
  \href{http://xxx.lanl.gov/abs/1903.06239}{{\tt 1903.06239}}.

\bibitem{ArkaniHamed:2006dz}
N.~Arkani-Hamed, L.~Motl, A.~Nicolis, and C.~Vafa, ``{The String landscape,
  black holes and gravity as the weakest force},'' {\em JHEP} {\bf 06} (2007)
  060, \href{http://xxx.lanl.gov/abs/hep-th/0601001}{{\tt hep-th/0601001}}.

\bibitem{Palti:2017elp}
E.~Palti, ``{The Weak Gravity Conjecture and Scalar Fields},'' {\em JHEP} {\bf
  08} (2017) 034, \href{http://xxx.lanl.gov/abs/1705.04328}{{\tt 1705.04328}}.

\bibitem{Ooguri:2006in}
H.~Ooguri and C.~Vafa, ``{On the Geometry of the String Landscape and the
  Swampland},'' {\em Nucl. Phys.} {\bf B766} (2007) 21--33,
  \href{http://xxx.lanl.gov/abs/hep-th/0605264}{{\tt hep-th/0605264}}.

\bibitem{Klaewer:2016kiy}
D.~Klaewer and E.~Palti, ``{Super-Planckian Spatial Field Variations and
  Quantum Gravity},'' {\em JHEP} {\bf 01} (2017) 088,
  \href{http://xxx.lanl.gov/abs/1610.00010}{{\tt 1610.00010}}.

\bibitem{Grimm:2018ohb}
T.~W. Grimm, E.~Palti, and I.~Valenzuela, ``{Infinite Distances in Field Space
  and Massless Towers of States},'' {\em JHEP} {\bf 08} (2018) 143,
  \href{http://xxx.lanl.gov/abs/1802.08264}{{\tt 1802.08264}}.

\bibitem{Grimm:2004uq}
T.~W. Grimm and J.~Louis, ``{The Effective action of N = 1 Calabi-Yau
  orientifolds},'' {\em Nucl. Phys.} {\bf B699} (2004) 387--426,
  \href{http://xxx.lanl.gov/abs/hep-th/0403067}{{\tt hep-th/0403067}}.

\bibitem{Font:2019cxq}
A.~Font, A.~Herr{\'a}ez, and L.~E. Ib{\'a}{\~n}ez, ``{The Swampland Distance
  Conjecture and Towers of Tensionless Branes},'' {\em JHEP} {\bf 08} (2019)
  044, \href{http://xxx.lanl.gov/abs/1904.05379}{{\tt 1904.05379}}.

\bibitem{martin:2019}
M.~Enr{\'i}quez, ``{Swampland Conjectures for $\mathcal{N}=1$ Orientifolds},''
  {\em
  \href{https://www.theorie.physik.uni-muenchen.de/TMP/theses/theisenriques.pdf}{M.Sc.~thesis}}
  (2019).
  \href{https://www.theorie.physik.uni-muenchen.de/TMP/theses/theisenriques.pdf}{[https://www.theorie.physik.uni-muenchen.de/TMP/theses]}.

\bibitem{Hosono:1994av}
S.~Hosono, A.~Klemm, and S.~Theisen, ``{Lectures on mirror symmetry},'' {\em
  Lect. Notes Phys.} {\bf 436} (1994) 235--280,
  \href{http://xxx.lanl.gov/abs/hep-th/9403096}{{\tt hep-th/9403096}}.

\bibitem{Hosono:1994ax}
S.~Hosono, A.~Klemm, S.~Theisen, and S.-T. Yau, ``{Mirror symmetry, mirror map
  and applications to complete intersection Calabi-Yau spaces},'' {\em Nucl.
  Phys.} {\bf B433} (1995) 501--554,
  \href{http://xxx.lanl.gov/abs/hep-th/9406055}{{\tt hep-th/9406055}}.
  [,545(1994); AMS/IP Stud. Adv. Math.1,545(1996)].

\bibitem{Strominger:1995cz}
A.~Strominger, ``{Massless black holes and conifolds in string theory},'' {\em
  Nucl. Phys.} {\bf B451} (1995) 96--108,
  \href{http://xxx.lanl.gov/abs/hep-th/9504090}{{\tt hep-th/9504090}}.

\bibitem{Jockers:2005pn}
H.~Jockers, ``{The Effective Action of D-branes in Calabi-Yau Orientifold
  Compactifications},'' {\em Fortsch. Phys.} {\bf 53} (2005) 1087--1175,
  \href{http://xxx.lanl.gov/abs/hep-th/0507042}{{\tt hep-th/0507042}}.

\bibitem{Marchesano:2014bia}
F.~Marchesano, D.~Regalado, and G.~Zoccarato, ``{U(1) mixing and D-brane linear
  equivalence},'' {\em JHEP} {\bf 08} (2014) 157,
  \href{http://xxx.lanl.gov/abs/1406.2729}{{\tt 1406.2729}}.

\bibitem{Bergshoeff:1996tu}
E.~Bergshoeff and P.~K. Townsend, ``{Super D-branes},'' {\em Nucl. Phys.} {\bf
  B490} (1997) 145--162, \href{http://xxx.lanl.gov/abs/hep-th/9611173}{{\tt
  hep-th/9611173}}.

\bibitem{Bergshoeff:1997kr}
E.~Bergshoeff, R.~Kallosh, T.~Ortin, and G.~Papadopoulos, ``{Kappa symmetry,
  supersymmetry and intersecting branes},'' {\em Nucl. Phys.} {\bf B502} (1997)
  149--169, \href{http://xxx.lanl.gov/abs/hep-th/9705040}{{\tt
  hep-th/9705040}}.

\bibitem{Marino:1999af}
M.~Marino, R.~Minasian, G.~W. Moore, and A.~Strominger, ``{Nonlinear instantons
  from supersymmetric p-branes},'' {\em JHEP} {\bf 01} (2000) 005,
  \href{http://xxx.lanl.gov/abs/hep-th/9911206}{{\tt hep-th/9911206}}.

\bibitem{Kachru:1999vj}
S.~Kachru and J.~McGreevy, ``{Supersymmetric three cycles and supersymmetry
  breaking},'' {\em Phys. Rev.} {\bf D61} (2000) 026001,
  \href{http://xxx.lanl.gov/abs/hep-th/9908135}{{\tt hep-th/9908135}}.

\bibitem{Lust:2006zh}
D.~L{\"u}st, S.~Reffert, E.~Scheidegger, and S.~Stieberger, ``{Resolved
  Toroidal Orbifolds and their Orientifolds},'' {\em Adv. Theor. Math. Phys.}
  {\bf 12} (2008), no.~1 67--183,
  \href{http://xxx.lanl.gov/abs/hep-th/0609014}{{\tt hep-th/0609014}}.

\bibitem{Becker:1995kb}
K.~Becker, M.~Becker, and A.~Strominger, ``{Five-branes, membranes and
  nonperturbative string theory},'' {\em Nucl. Phys.} {\bf B456} (1995)
  130--152, \href{http://xxx.lanl.gov/abs/hep-th/9507158}{{\tt
  hep-th/9507158}}.

\bibitem{Sen:1999mg}
A.~Sen, ``{NonBPS states and Branes in string theory},'' in {\em {Supersymmetry
  in the theories of fields, strings and branes. Proceedings, Advanced School,
  Santiago de Compostela, Spain, July 26-31, 1999}}, pp.~187--234, 1999.
\newblock \href{http://xxx.lanl.gov/abs/hep-th/9904207}{{\tt hep-th/9904207}}.
\newblock [,45(1999)].

\bibitem{Sen:1999nx}
A.~Sen and B.~Zwiebach, ``{Tachyon condensation in string field theory},'' {\em
  JHEP} {\bf 03} (2000) 002, \href{http://xxx.lanl.gov/abs/hep-th/9912249}{{\tt
  hep-th/9912249}}.

\bibitem{Heidenreich:2017sim}
B.~Heidenreich, M.~Reece, and T.~Rudelius, ``{The Weak Gravity Conjecture and
  Emergence from an Ultraviolet Cutoff},'' {\em Eur. Phys. J.} {\bf C78}
  (2018), no.~4 337, \href{http://xxx.lanl.gov/abs/1712.01868}{{\tt
  1712.01868}}.

\bibitem{Heidenreich:2019zkl}
B.~Heidenreich, M.~Reece, and T.~Rudelius, ``{Repulsive Forces and the Weak
  Gravity Conjecture},'' {\em JHEP} {\bf 10} (2019) 055,
  \href{http://xxx.lanl.gov/abs/1906.02206}{{\tt 1906.02206}}.

\bibitem{Polchinski:2003bq}
J.~Polchinski, ``{Monopoles, duality, and string theory},'' {\em Int. J. Mod.
  Phys.} {\bf A19S1} (2004) 145--156,
  \href{http://xxx.lanl.gov/abs/hep-th/0304042}{{\tt hep-th/0304042}}.
  [,145(2003)].

\bibitem{Heidenreich:2015nta}
B.~Heidenreich, M.~Reece, and T.~Rudelius, ``{Sharpening the Weak Gravity
  Conjecture with Dimensional Reduction},'' {\em JHEP} {\bf 02} (2016) 140,
  \href{http://xxx.lanl.gov/abs/1509.06374}{{\tt 1509.06374}}.

\bibitem{Heidenreich:2016aqi}
B.~Heidenreich, M.~Reece, and T.~Rudelius, ``{Evidence for a sublattice weak
  gravity conjecture},'' {\em JHEP} {\bf 08} (2017) 025,
  \href{http://xxx.lanl.gov/abs/1606.08437}{{\tt 1606.08437}}.

\bibitem{Andriolo:2018lvp}
S.~Andriolo, D.~Junghans, T.~Noumi, and G.~Shiu, ``{A Tower Weak Gravity
  Conjecture from Infrared Consistency},'' {\em Fortsch. Phys.} {\bf 66}
  (2018), no.~5 1800020, \href{http://xxx.lanl.gov/abs/1802.04287}{{\tt
  1802.04287}}.

\bibitem{Gaiotto:2014kfa}
D.~Gaiotto, A.~Kapustin, N.~Seiberg, and B.~Willett, ``{Generalized Global
  Symmetries},'' {\em JHEP} {\bf 02} (2015) 172,
  \href{http://xxx.lanl.gov/abs/1412.5148}{{\tt 1412.5148}}.

\bibitem{Dvali:2007hz}
G.~Dvali, ``{Black Holes and Large N Species Solution to the Hierarchy
  Problem},'' {\em Fortsch. Phys.} {\bf 58} (2010) 528--536,
  \href{http://xxx.lanl.gov/abs/0706.2050}{{\tt 0706.2050}}.

\bibitem{Haack:2017vko}
M.~Haack and J.~U. Kang, ``{Towards the 1-loop effective action of type IIB
  orientifolds},'' {\em PoS} {\bf CORFU2016} (2017) 099.

\bibitem{Blumenhagen:2019vgj}
R.~Blumenhagen, M.~Brinkmann, and A.~Makridou, ``{Quantum Log-Corrections to
  Swampland Conjectures},'' \href{http://xxx.lanl.gov/abs/1910.10185}{{\tt
  1910.10185}}.

\bibitem{Ooguri:2018wrx}
H.~Ooguri, E.~Palti, G.~Shiu, and C.~Vafa, ``{Distance and de Sitter
  Conjectures on the Swampland},'' {\em Phys. Lett.} {\bf B788} (2019)
  180--184, \href{http://xxx.lanl.gov/abs/1810.05506}{{\tt 1810.05506}}.

\bibitem{Grimm:2004ua}
T.~W. Grimm and J.~Louis, ``{The Effective action of type IIA Calabi-Yau
  orientifolds},'' {\em Nucl. Phys.} {\bf B718} (2005) 153--202,
  \href{http://xxx.lanl.gov/abs/hep-th/0412277}{{\tt hep-th/0412277}}.

\bibitem{Corvilain:2018lgw}
P.~Corvilain, T.~W. Grimm, and I.~Valenzuela, ``{The Swampland Distance
  Conjecture for Kähler moduli},'' {\em JHEP} {\bf 08} (2019) 075,
  \href{http://xxx.lanl.gov/abs/1812.07548}{{\tt 1812.07548}}.

\bibitem{vittmann:2019}
C.~Vittmann, ``{The Axion-Instanton Weak Gravity Conjecture and Scalar
  Fields},'' {\em
  \href{https://wwwth.mpp.mpg.de/members/palti/images/ClemensThesis.pdf}{M.Sc.~thesis}}
  (2018).
  \href{https://wwwth.mpp.mpg.de/members/palti/images/ClemensThesis.pdf}{[https://wwwth.mpp.mpg.de/members/palti/images/ClemensThesis.pdf]}.

\bibitem{DallAgata:2020ino}
G.~Dall'Agata and M.~Morittu, ``{Covariant formulation of BPS Black holes and
  the scalar weak gravity conjecture},''
  \href{http://xxx.lanl.gov/abs/2001.10542}{{\tt 2001.10542}}.

\end{thebibliography}\endgroup
\bibliographystyle{utphys}


\end{document}